\documentclass[twoside]{article}
\usepackage[a4paper]{geometry}
\usepackage[latin1]{inputenc} % ou \usepackage[utf8]{inputenc}
\usepackage[T1]{fontenc} % ou \usepackage[OT1]{fontenc}
\usepackage{RR}
\usepackage{hyperref}

\usepackage{amsmath}
\usepackage{amssymb}
\usepackage{todonotes}
\usepackage{bbm}

\newtheorem{theorem}{Theorem}

\newtheorem{defi}{Definition}

\newcommand{\ones}{\underline{1}}
\newcommand{\qed}{\hfill $\Box$}

\newcommand{\prob}{{\mathbb P}}
\newcommand{\expec}{{\mathbb E}}
\newcommand{\eqn}[1]{\begin{equation} #1 \end{equation}}
\newcommand{\eqan}[1]{\begin{align} #1 \end{align}}
\newcommand{\sss}{\scriptscriptstyle}

\newcommand{\nn}{\nonumber}

\newcommand\1{\mathbbm{1}}
\newcommand{\indic}[1]{\1_{\{#1\}}}

\newcommand{\Bin}{{\sf Bin}}

\newcommand {\convas}{\stackrel{\sss a.s.}{\longrightarrow}}

\newcommand{\proof} {\noindent {\bf Proof}. \hspace{2mm}}

\RRdate{July 2014}
\RRauthor{% les auteurs
Konstantin Avrachenkov\thanks{Inria Sophia Antipolis, France, {\tt k.avrachenkov@sophia.inria.fr}}%
  \and
Remco van der Hofstad\thanks{Eindhoven University of Technology,
The Netherlands, {\tt r.w.v.d.hofstad@tue.nl}}  
\and \\
Marina Sokol\thanks{Inria Sophia Antipolis, France, {\tt marina.sokol@inria.fr}}
}

\authorhead{K. Avrachenkov \& R. W. v. d. Hofstad \& M. Sokol}
\RRtitle{PageRank Personnalis\'e avec la Probabilit\'e d'un Red\'emarrage
en Fonction de N{\oe}ud}
\RRetitle{Personalized PageRank with Node-dependent Restart}
\titlehead{Personalized PageRank with Node-dependent Restart}
\RRresume{
PageRank personnalis\'e est un algorithme permettant de classer les pages web par l'importance pertinente \`a l'utilisateur.
Nous introduisons deux g\'en\'eralisations de PageRank personnalis\'e avec la probabilit\'e d'un red\'emarrage
en fonction de n{\oe}ud. La premi\`ere g\'en\'eralisation est bas\'ee sur la proportion de visites aux n{\oe}uds avant 
le red\'emarrage, tandis que la seconde g\'en\'eralisation est bas\'ee sur la probabilit\'e de la visite juste avant le red\'emarrage. 
Dans le cas original de PageRank personnalis\'e, la probabilit\'e de red\'emarrage est constante et les deux nouvelles mesures 
co\"{\i}ncident. Nous discutons des cas particuliers int\'eressants de la probabilit\'e de red\'emarrage et la distribution
de red\'emarrage. Nous montrons que les deux g\'en\'eralisations de PageRank personnalis\'e ont des expressions \'el\'egantes 
reliant les "directe" et "inverse" PageRanks personnalis\'es.
}
\RRabstract{
Personalized PageRank is an algorithm to classify the improtance of web pages on a user-dependent basis. We introduce two generalizations of Personalized PageRank with node-dependent restart. The first generalization is based on the proportion of visits to nodes before the restart, whereas the second generalization is based on the probability of visited node just before the restart. In the original case of constant restart probability, the two measures coincide. We discuss interesting particular cases of restart probabilities and restart distributions. We show that the both generalizations of Personalized PageRank have an elegant expression connecting the so-called direct and reverse Personalized PageRanks that yield a symmetry property of these Personalized PageRanks.
}
\RRmotcle{PageRank, Red\'emarrage en Fonction de N{\oe}ud, Marche Al\'eatoire sur un Graphe}

\RRkeyword{PageRank, Node-dependant Restart Probability, Random Walk on Graph}

\RRprojets{Maestro}
\URSophia % pour ceux qui sont au Sud.
\begin{document}
\RRNo{8570}
\makeRR   % cas d'un rapport de recherche

\section{Introduction and definitions}
PageRank has become a standard algorithm to classify the importance of nodes in a network. Let us start by introducing some notation. Let $G=(V,E)$ be a finite graph, where $V$ is the node set and $E\subseteq V\times V$ the collection of (directed) edges. Then, PageRank can be interpreted as the stationary distribution of a random walk on $G$ that restarts from a uniform location in $V$ at each time with probability $\alpha\in (0,1)$. Thus, in the Standard PageRank centrality measure \cite{BPMW98}, the random walk restarts after a geometrically distributed number of steps, and the restart takes place from a uniform location in the graph, and otherwise jumps to any one of the neighbours in the graph with equal probability. Personalized PageRank \cite{H02} is a modification of the Standard PageRank where the restart distribution is not uniform. Both the Standard and Personalized PageRank have many applications in data mining and machine learning (see e.g., \cite{ADNPS08,AGMS12,BPMW98,CXMR07,Fetal12,H02,LBNS05,MA07}).

In the (standard) Personalized PageRank, the random walker restarts with a given fixed probability $1-\alpha$ at each visited node. We suggest a generalization where a random walker restarts with probability $1-\alpha_i$ at node $i\in V$. When the random walker restarts, it chooses a node to restart at with probability distribution $v^T$. In many cases, we let the random walker restart at a fixed location, say $j\in V$. Then the Personalized PageRank of node $j$ corresponds to $j$th Personalized PageRank and is a vector whose $i$th coordinate measures the importance of node $i$ to node $j$.

The above random walks $(X_t)_{t\geq 0}$ can be described by a finite-state Markov chain with the transition matrix
	\begin{equation}
	\label{eq:PPRAtrans}
	\tilde {P} = A D^{-1} W + (I-A) \ones v^T,
	\end{equation}
where $W$ is the (possibly non-symmetric) adjacency matrix, $D$ is the diagonal matrix with diagonal entries $D_{ii}=\sum_{j=1}^n W_{ij}$, and $A=\mbox{diag} (\alpha_1,\ldots,\alpha_n)$ is the diagonal matrix of damping factors. The case of undirected graphs corresponds to the case when $W$ is a symmetric matrix. In general, $D_{ii}$ is the out-degree of node $i\in V$. Throughout the paper, we assume that the graph is
weakly connected and if some node does not have outgoing edges, we add artificial outgoing edges to all the other nodes.

We propose two generalizations of the Personalized PageRank with node-dependent
restart:

\begin{defi}[Occupation-time Personalized PageRank]
\label{def-OtPPR}
The \emph{Occupation-Time Personalized PageRank} is given by
	\eqn{
	\pi_j(v) = \lim_{t\rightarrow \infty} \prob(X_t=j).
	}
\end{defi}

\noindent
By the fact that $(\pi_j(v))_{v\in V}$ is the stationairy distribution of the Markov chain, we can interpret $\pi_j(v)$ as a long-run frequency of visits to node $j$, i.e.,
	\eqn{
	\pi_j(v)=\lim_{t\rightarrow \infty} \frac{1}{t} \sum_{s=1}^t \indic{X_s=v}.
	}

Our second generalization is based on the location where the random walker restarts:

\begin{defi}[Location-of-Restart Personalized PageRank]
\label{def-LoRPPR}
The \emph{Location-of-Restart Personalized PageRank} is given by
	\eqn{
	\rho_j(v) = \lim_{t\rightarrow \infty} \prob(X_t=j \ \mbox{\rm just before restart})
	=\lim_{t\rightarrow \infty} \prob(X_t=j \mid \mbox{\rm restart at time }t+1).
	}
\end{defi}

\noindent We can interpret $\rho_j(v)$ as a long-run frequency of visits to
node $j$ which are followed immediately by a restart, i.e.,
	\eqn{
	\rho_j(v) = \lim_{t\rightarrow \infty}
	\frac{1}{N_t}\sum_{s=1}^t \indic{X_t=j, X_{t+1} \text{ restarts}},
	}
where $N_t$ denotes the number of restarts up to time $t$. When the restarts occur with equal probability for every node, we have that $N_t\sim \Bin(t,1-\alpha)$, i.e., $N_t$ has a binomial distribution with $t$ trials and success probability $1-\alpha$. When the restart probabilities are unequal, the distribution of $N_t$ is more involved. In general, however,
	\eqn{\label{eq:restartsfrac}
	N_t/t \convas \sum_{j\in V} (1-\alpha_j) \pi_j(v),
	}
where $\convas$ denotes convergence almost surely.

Both generalized Personalized PageRanks are probability distributions, i.e., their sum over $j\in V$ gives 1.
When $v^T=e(i)$, where $e_j(i)=1$ when $i=j$ and $e_j(i)=0$ when $i\neq j$, then both $\pi_j(v)$ and $\rho_j(v)$ can be interpreted as the relative importance of node $j$ from the perspective of node $i$.

We see at least three applications of the generalized Personalized PageRank. The network sampling process introduced in \cite{ART10} can be viewed as a particular case of PageRank with a node-dependent restart. We discuss this relation in more detail in Section~\ref{sec-special}. Secondly, the generalized Personalized PageRank can be applied as a proximity measure between nodes in semi-supervised machine learning \cite{AGS13,Fetal12}. In this case, one may prefer to discount the effect of less informative nodes, e.g., nodes with very large degrees. And thirdly, the generalized Personalized PageRank can be applied for spam detection and control. It is known \cite{Cetal07} that spam web pages are often designed to be ranked highly. By using the Location-of-Restart Personalized PageRank and penalizing the ranking of spam pages with small restart probability, one can push the spam pages from the top list produced by search engines.

In this paper, we investigate these two generalizations of Personalized PageRank.
The paper is organised as follows. In Section \ref{sec-OtPR}, we investigate the Occupation-Time Personalized PageRank. In Section \ref{sec-LoRPR}, we investigate the Location-of-Restart Personalized PageRank. In Section \ref{sec-special}, we specify the results for some particular interesting cases. We close in Section \ref{sec-disc} with a discussion of our results and suggestions for future research.

%In Section \ref{sec-mon-prop}, we investigate monotonicity properties of the our Personalized PageRank
%measures with respect to $\alpha_j$.

\section{Occupation-time Personalized PageRank}
\label{sec-OtPR}

The Occupation-time Personalized PageRank can be calculated explicitly as follows:

\begin{theorem}[Occupation-time Personalized PageRank Formula]
\label{thm-OtPR}
The Occupation-time Personalized PageRank $\pi(v)$ with node-dependent restart
equals
	\begin{equation}
	\label{eq:PPRAcf}
	\pi(v) = \frac{1}{v^T [I-AP]^{-1} \ones} v^T [I- AP]^{-1},
	\end{equation}
with $P=D^{-1}W$ the transition matrix of random walk on $G$ withour restarts.
\end{theorem}

\proof
By the defining equation for the stationary distribution of a Markov chain,
	\eqn{
	\pi(v) [A D^{-1} W + (I-A) \ones v^T] = \pi(v),
	}
so that
	\eqn{
	\pi(v) [I - A D^{-1} W] = \pi(v) (I-A) \ones v^T,
	}
and, since $\pi(v) \ones=1$,
	\eqn{
	\pi(v) [I - A D^{-1} W] = (1- \pi(v) A \ones) v^T.
	}
Since the matrix $A D^{-1} W$ is substochastic and hence $[I - A D^{-1} W]$ is invertible,
we arrive at
	\eqn{
	\pi(v) = (1- \pi(v) A \ones) v^T [I - A D^{-1} W]^{-1}.
	}
Let us multiply the above equation from the right hand side by $A\ones$ to obtain
	\eqn{
	\pi(v)A\ones = (1- \pi(v) A \ones) v^T [I - A D^{-1} W]^{-1} A \ones.
	}
This yields
	\eqn{
	\pi(v)A\ones = \frac{v^T [I-AP]^{-1}A \ones}{1 + v^T [I-AP]^{-1}A \ones},
	}
and, consequently, since $A=\mbox{diag} (\alpha_1,...,\alpha_n)$ is a diagonal matrix,
so that $A \ones=(\alpha_1,...,\alpha_n)^T$, and we arrive at
	\eqn{
	\pi(v) = \frac{1}{1+v^T [I-AP]^{-1}A \ones} v^T [I- AP]^{-1}.
	}
Since $v^T\ones=1$, by the fact that $v^T$ is a probability mass function, we obtain
	\eqn{
	1+v^T [I-AP]^{-1}A \ones = v^T [I-AP]^{-1} \ones,
	}
from which the required equation (\ref{eq:PPRAcf}) follows. \qed

\bigskip
Formula (\ref{eq:PPRAcf}) admits the following probabilistic interpretation
in the form of renewal equation
	\eqn{
	\label{prob-OtPR}
	\pi_j(v) = \frac{\expec_v[\mbox{\rm \# visits to $j$ before restart}]}{\expec_v[\mbox{\rm \# steps before restart}]},
	}
where $\expec_v$ denotes expectation with respect to the Markov chain starting in distribution $v$.

Denote for brevity $\pi_j(i)=\pi_j(e_i^T)$, where $e_i$ is the $i$th vector of the standard basis, so that
$\pi_j(i)$ denotes the importance of node $j$ from the perspective of $i$. Similarly, $\pi_i(j)$ denotes the importance of node $i$ from the perspective of $j$. We next prove a relation between these ``direct'' and ``reverse'' PageRanks in the case of \emph{undirected} graphs.

\begin{theorem}[Symmetry for undirected Occupation-time Personalized PageRank]
\label{thm-symmOtPPRu}
When $W^T=W$ and $A>0$, the following relation holds
	\begin{equation}
	\label{eq:dirrev}
	\frac{d_i}{\alpha_i K_i(A)} \pi_j(i) = \frac{d_j}{\alpha_j K_j(A)}\pi_i(j),
	\end{equation}
with
	\eqn{
	K_i(A) = \frac{1}{e_i^T [I-AP]^{-1} \ones}.
	}
\end{theorem}

\proof Note that the denominator of \eqref{eq:PPRAcf} equals precisely $K_i(A)$. Thus, using a matrix geometric series expansion, we can rewrite equation \eqref{eq:PPRAcf} as
	\eqan{
	\pi_j(i) 	&= K_i(A) e_i^T \sum_{k=0}^\infty (AD^{-1}W)^k e_j\\
			&= K_i(A) e_i^T \sum_{k=0}^\infty (AD^{-1}W)^k D^{-1} A A^{-1} D e_j\nn\\
			&= K_i(A) e_i^T A D^{-1} \sum_{k=0}^\infty (WD^{-1}A)^k A^{-1} D e_j\nn\\
			&= K_i(A) \frac{\alpha_i}{d_i} e_i^T \sum_{k=0}^\infty (WD^{-1}A)^k e_j \frac{d_j}{\alpha_j}\nn\\
			&= \frac{K_i(A)}{K_j(A)} \frac{\alpha_i}{d_i} \frac{d_j}{\alpha_j} K_j(A) e_i^T [I - WD^{-1}A]^{-1} e_j\nn\\
			&= \frac{K_i(A)}{K_j(A)} \frac{\alpha_i}{d_i} \frac{d_j}{\alpha_j} K_j(A) e_j^T [I - AD^{-1}W]^{-1} e_i,\nn
}
which gives equation (\ref{eq:dirrev}).
\qed

\bigskip

We note that the term $(AD^{-1}W)^k$ can be interpreted as the contribution corresponding
to all paths of length $k$, while $K_i(A)$ can be interpreted as the reciprocal of the expected
time between two consecutive restarts if the restart distribution is concentrated on node $i$, i.e.,
	\eqn{
	K_i(A)^{-1}=\expec_i[\mbox{\rm \# steps before restart}],
	}
see also \eqref{prob-OtPR}. Thus, a probabilistic interpretation of
\eqref{eq:PPRAcf} is that
	\eqn{
	\label{prob-OtPR}
	\frac{d_i}{\alpha_i}\expec_i[\mbox{\rm \# visits to $j$ before restart}]
	=\frac{d_j}{\alpha_j}\expec_j[\mbox{\rm \# visits to $i$ before restart}].
	}
Since
	\eqn{
    \label{Enumvisits}
	\expec_i[\mbox{\rm \# visits to $j$ before restart}]
	=\sum_{k=1}^{\infty} \sum_{v_1, \ldots, v_k} \prod_{t=0}^{k-1} \frac{\alpha_{v_s}}{d_{v_s}},
	}
where $v_0=j$, we immediately see that the expression for $\expec_j[\mbox{\rm \# visits to $i$ before restart}]$ is identical, except for the first factor of $\frac{\alpha_i}{d_i}$, which is present in
$\expec_i[\mbox{\rm \# visits to $j$ before restart}]$, but not in $\expec_i[\mbox{\rm \# visits to $j$ before restart}]$, and the factor $\frac{\alpha_j}{d_j}$, which is present in
$\expec_j[\mbox{\rm \# visits to $i$ before restart}]$, but not in $\expec_j[\mbox{\rm \# visits to $i$ before restart}]$. This explains the factors $\frac{d_i}{\alpha_i}$ and $\frac{d_j}{\alpha_j}$ in \eqref{prob-OtPR} and gives an alternative probabilistic proof of Theorem \ref{thm-symmOtPPRu}.

\section{Location-of-Restart Personalized PageRank}
\label{sec-LoRPR}
The Location-of-Restart Personalized PageRank can also be calculated explicitly:

\begin{theorem}[Location-of-Restart Personalized PageRank Formula]
\label{thm-LoRPR}
The Location-of-Restart Personalized PageRank $\rho(v)$ with node-dependent restart
is equal to
	\begin{equation}
	\label{eq:PPRBcf}
	\rho(v) =  v^T [I- AP]^{-1} [I-A],
	\end{equation}
with $P=D^{-1}W$.
\end{theorem}

\proof This follows from the formula
	\eqan{
	\label{LoRPPR}
	\rho_j(v) &= \expec_v[\mbox{\rm \# visits to $j$ before restart}] \prob(\mbox{\rm restart from $j$})\\
	&= \expec_v[\mbox{\rm \# visits to $j$ before restart}] (1-\alpha_j).\nn
	}
Now we can use \eqref{Enumvisits} and the analysis in the proof of Theorem \ref{thm-OtPR} to complete the proof.
%\vskip-0.8cm
\qed
\vskip0.8cm

Location-of-Restart Personalized PageRank admits an even more
elegant relation between the ``direct'' and ``reverse'' PageRanks in the
case of undirected graphs:

\begin{theorem}[Symmetry for undirected Location-of-Restart Personalized PageRank]
\label{thm-symmLoRPPRu}
When $W^T=W$ and $\alpha_i \in (0,1)$, the following relation holds
\begin{equation}
\label{eq:dirrev2}
\frac{1-\alpha_i}{\alpha_i} \ d_i \ \rho_j(i) = \frac{1-\alpha_j}{\alpha_j} \ d_j \ \rho_i(j).
\end{equation}
\end{theorem}
\proof This follows from a series of equivalent transformations
	\eqan{
	\rho_j(i) 	&= e_i^T [I-AP]^{-1} [I-A] e_j = e_i^T [I-AP]^{-1} e_j (1-\alpha_j)\\
			&= e_i^T [AD^{-1}(DA^{-1}-W)]^{-1} e_j (1-\alpha_j)
				= e_i^T [DA^{-1}-W]^{-1} e_j d_j \frac{1-\alpha_j}{\alpha_j}\nn\\
			&= e_i^T [(I-WD^{-1}A)DA^{-1}]^{-1} e_j d_j \frac{1-\alpha_j}{\alpha_j}
				= e_i^T AD^{-1} [I-WD^{-1}A]^{-1} e_j d_j \frac{1-\alpha_j}{\alpha_j}\nn\\
			&= \frac{\alpha_i}{d_i} e_i^T [I-WD^{-1}A]^{-1} e_j d_j \frac{1-\alpha_j}{\alpha_j}\nn\\
			&= \frac{\alpha_i}{d_i} \frac{\rho_i(j)}{1-\alpha_i} d_j \frac{1-\alpha_j}{\alpha_j}.\nn
	}
Alternatively, Theorem \ref{thm-symmLoRPPRu} follows directly from \eqref{LoRPPR} and \eqref{prob-OtPR}.
\qed
\medskip

Interestingly, in (\ref{eq:dirrev}), the whole graph topology
has an effect on the relation between the ``direct'' and ``reverse'' Personalized PageRanks, whereas in
the case of $\rho(v)$, see equation (\ref{eq:dirrev2}), only the local end-point information
(i.e., $\alpha_i$ and $d_i$) have an effect on the relation between the ``direct'' and ``reverse'' PageRanks.
We have no intuitive explanation of this distinction.

%\section{Monotonicity properties with respect to restart probabilities}
%\label{sec-mon-prop}
%\KA{Are there some monotonicity properties with respect to $\alpha_j$?}
%\RvdH{Need to think about this...}

\section{Interesting particular cases}
\label{sec-special}
In this section, we consider some interesting particular cases for the choice of restart probabilities
and distributions.

\subsection{Constant probability of restart}
\label{sec-constant-restart}
The case of constant restart probabilities (i.e., $\alpha_j=\alpha$ for every $j$) corresponds to the
original or standard Personalized PageRank. We note that in this case the two generalizations coincide. For instance,  we can recover a known formula \cite{moler} for the original Personalized PageRank with $A=\alpha I$ from equation (\ref{eq:PPRAcf}). Specifically,
	\eqn{
	v^T [I-AP]^{-1} \ones = \alpha v^T [I - \alpha P]^{-1} \ones =
	v^T \sum_{k=0}^\infty \alpha^k P^k \ones = \frac{1}{1-\alpha},
	}
and hence we retrieve the well-known formula
	\begin{equation}
	\label{eq:PPRclas}
	\pi(v) = (1-\alpha) v^T [I- \alpha P]^{-1}.
	\end{equation}

We also retrieve the following elegant result connecting direct and ``reverse''
original Personalized PageRanks on undirected graphs ($W^T=W$) obtained in \cite{AGS13}:
	\begin{equation}
	\label{eq:dirrevclass}
	d_i \pi_j(i) = d_j \pi_i(j),
	\end{equation}
since in the original Personalized PageRank $\alpha_i=\alpha$. Finally, we note that
in the original Personalized PageRank, the expected time between restart does not
depend on the graph structure nor on the restart distribution and is given by
	\eqn{
	\expec_v[\mbox{\rm time between consecutive restarts}] = \frac{1}{1-\alpha},
	}
which is just the mean of the geomatrically distributed random variable.

\subsection{Restart probabilities proportional to powers of degrees}
\label{sec-restart-degrees}

Let us consider a particular case when the restart probabilities are proportional
to powers of the degrees. Namely, let
	\eqn{
	\label{A-form-degree}
	A = I - a D^\sigma,
	}
with $a d^\sigma_{{\mathrm {max}}} < 1$. We first analyse $[I -AP]^{-1}$ with the help of a Laurent series expansion. Let $T(\varepsilon)=T_0-\varepsilon T_1$ be
a substochastic matrix for small values of $\varepsilon$ and let $T_0$ be a stochastic matrix with associated stationary distribution $\xi^T$ and deviation matrix $H=(I-T_0+\ones \xi^T)^{-1}-\ones \xi^T$.   Then, the following Laurent series expansion takes place (see Lemma~6.8 from \cite{AFH13})
	\eqn{
	[I-T(\varepsilon)]^{-1} = \frac{1}{\varepsilon} X_{-1} +
	X_{0} + \varepsilon X_{1} +\ldots,
	}
where the first two coefficients are given by
	\eqn{
	\label{eq:X-1}
	X_{-1} = \frac{1}{\pi^T T_1 \ones} \ones \xi^T,
	}
and
	\eqn{
	\label{eq:X0}
	X_{0} = (I-X_{-1}T_1)H(I-T_1X_{-1}).
	}
Applying the above Laurent power series to $[I-AP]^{-1}$ with $T_0=P$, $T_1=D^\sigma P$ and	$\varepsilon=a$, we obtain
	\eqn{
	\label{eq:mainterm}
	[I-AP]^{-1} = [I-(P-aD^\sigma P)]^{-1}
	= \frac{1}{a} \frac{1}{\pi^T T_1 \ones} \ones \xi^T + \mbox{O}(a)
	= \frac{1}{a} \frac{1}{\xi^T D^\sigma \ones} \ones \xi^T + \mbox{O}(a).
	}
This yields the following asymptotic expressions for the generlized Personalized PageRanks
	\eqn{
	\label{pipropdeg}
	\pi_j(a) = \xi_j + \mbox{o}(a),
	}
and
	\eqn{
	\label{pipropdeg}
	\rho_j(a) = \frac{d_j^\sigma \xi_j}{\sum_{i\in V} d_i^\sigma \xi_i} + \mbox{o}(a).
	}
In particular, if we assume that the graph is undirected ($W^T=W$), we can further specify the above expressions
	\eqn{
	\pi_j(a) = \frac{d_j}{\sum_i d_i} + \mbox{o}(a),
	}
and
	\eqn{
	\rho_j(a) = \frac{d_j^{1+\sigma}}{\sum_{i\in V} d_i^{1+\sigma}} + \mbox{o}(a).
	}
We observe that using positive or negative degree $\sigma$ we can significantly
penalize or promote the score $\rho$ for nodes with large degrees.

As a by-product of our computations, we have also obtain nice asymptotic expression
for the expected time between restarts in the case of undirected graph:
	\eqn{
	\expec_v[\mbox{\rm time between consecutive restarts}]
	= \frac{1}{a} \frac{\sum_{i\in V} d_i}{\sum_{i\in V} d_i^{1+\sigma}}
	+ \mbox{O}(a).
	}
One interesting conclusion from the above expression is that when $\sigma>0$ the highly skewed distribution of the degree distribution in $G$ can significantly shorten the time between restarts.

\subsection{Random walk with jumps}
\label{sec-special-jumps}
In \cite{ART10}, the authors introduced a process with artificial jumps. It is suggested
in \cite{ART10} to add artificial edges with weights $a/n$ between each two
nodes to the graph. This process creates self-loops as well. Thus, the new modified graph is a combination of the original graph and a complete graph with self-loops. Let us demonstrate that this is a particular case
of the introduce generalized definition of Personalized PageRank. Specifically, we define
the damping factors as
	\eqn{\label{eq:alpha}
	\alpha_i = \frac{d_i}{d_i+a}, \quad i\in V,
	}
and as the restart distribution we take the uniform distribution ($v = \ones/n$).
Indeed, it is easy to check that we retrieve the transition probabilities from \cite{ART10}
	\eqn{
	p_{ij} = \left\{ \begin{array}{ll}
	\frac{a+n}{n(d_i+a)} &\qquad\mbox{when $i$ has an edge to $j$},\\
	\frac{a}{n(d_i+a)} &\qquad \mbox{when $i$ does not have an edge to $j$}.
	\end{array}\right.
	}
As was shown in \cite{ART10}, the stationary distribution of the modified process,
coinciding with the Occupation-time Personalized PageRank, is given by
	\begin{equation}
	\label{eq:piex3}
	\pi_i = \pi_i(\ones/n)=\frac{d_i+a}{2|E|+na}, \quad i\in V.
	\end{equation}
In particular, from (\ref{eq:restartsfrac}) we conclude that in the stationary regime
$$
\expec_v[\mbox{\rm time between consecutive restarts}]
= \left( \sum_{j \in V} \left(1-\frac{d_j}{d_j+a}\right) \frac{d_j+a}{2|E|+na} \right)^{-1}
$$
$$
= \frac{2|E|+na}{na} 
= \frac{\bar{d}+a}{a},
$$
where $\bar{d}$ is the average degree of the graph.
Since $\pi(v)$ is the stationary distribution of $\tilde{P}$ with $v = \ones/n$ (see (\ref{eq:PPRAtrans})),
it satisfies the equation
	\eqn{
	\pi (AP +[I-A]\ones v^T) = \pi.
	}
Rewriting this equation as
	\eqn{
	\pi [I-A] \ones v^T = \pi [I - AP],
	}
and postmultiplying by $[I - AP]^{-1}$, we obtain
	\eqn{
	\pi [I-A] \ones v^T [I - AP]^{-1} = \pi
	}
or
	\eqn{
	v^T [I - AP]^{-1} = \frac{\pi}{\sum_{i=1}^n \pi_i(1-\alpha_i)}.
	}
This yields
	\begin{equation}
	\label{eq:relatrhopi}
	\rho_j(v) = \frac{\pi_j(1-\alpha_j)}{\sum_{i=1}^n \pi_i(1-\alpha_i)}.
	\end{equation}
In our particular case of $\alpha_i=d_i/(d_i+a)$, the combination
of (\ref{eq:piex3}) and (\ref{eq:relatrhopi}) gives that $\pi_j(1-\alpha_j)$ is independent of $j$, so that
	\eqn{\label{eq:RWJrho}
	\rho_j = 1/n.
	}
This is quite surprising. Since $v^T=\frac{1}{n}\ones^T$, the nodes
just after restart are distributed uniformly. However, it appears
that the nodes just before restart are also uniformly distributed!
Such effect has also been observed in \cite{ALST13}.
Algorithmically, this means that all pages receive the {\em same} generalized Personalized PageRank $\rho$,  which, for ranking purposes, is rather uninformative. On the other hand, this Personalized PageRank can be useful for sampling procedures.
In fact, we can generalize (\ref{eq:alpha}) to
	\eqn{\label{eq:alphagen}
	\alpha_i = \frac{d_i}{d_i+a_i}, \quad i\in V,
	}
where now each node has its own parameter $a_i$. Now it is convenient to take as the restart
distribution
$$
v_i = \frac{a_i}{\sum_{k \in V}a_k}.
$$ 
Performing similar calculations as above, we arrive at
$$
\pi_j(v) = \frac{d_j+a_j}{2|E| + \sum_{k \in V}a_k}, \quad i\in V,
$$
and
$$
\rho_j(v) = \frac{a_i}{\sum_{k \in V}a_k}, \quad i\in V.
$$
Now in contrast with (\ref{eq:RWJrho}), the Location-of-Restart Personalized PageRank can be tuned.

\section{Discussion}
\label{sec-disc}

We have proposed two generalizations of Personalized PageRank when the probability of restart depends on the node. Both generalizations coincide with the original Personalized PageRank when the probability of restart is the same for all nodes. However, in general they show quite different behavior. In particular, the Location-of-Restart Personalized Pagerank appears to be stronger affected by the value of the restart probabilities. We have further suggested several applications of the generalized Personalized PageRank in machine learning, sampling and information retrieval and analized some particular interesting cases.

We feel that the analysis of the generalized Personalized PageRank on random graph model is a promising future research directions. We have already obtained some indications that the degree distribution can strongly affect the time between restarts. It would be highly interesting to analyse this effect in more detail on various random graph models (see e.g., \cite{Hofs14} for a introduction into random graphs, and \cite{CheOlv12} for first results on directed configuration models).

%
%\RvdH{Add discussion, also about the fact that power-laws of in-degrees and PageRank are empirically similar.
%How is this affected by the Personalized PageRanks that we choose?}

\paragraph{Acknowledgements.}
The work of KA and MS was partially supported by the EU project Congas and Alcatel-Lucent Inria Joint Lab.
The work of RvdH was supported in part by Netherlands
Organisation for Scientific Research (NWO). This work was initiated during the
`Workshop on Modern Random Graphs and Applications' held at Yandex, Moscow, October 24-26, 2013.
We thank Yandex, and in particular Andrei Raigorodskii, for bringing KA and RvdH together in such a wonderful setting.

\tableofcontents

\end{document}